# High-$T_c$ via electron polar coupling: relation to low-$T_c$ superconductivity and to chiral symmetry in particle physics

B. A. Green[a]
M. D. Scadron[b]

[a]11912 N. Centaurus Pl., Tucson AZ 85727
[b]Physics Department, University of Arizona, Tucson AZ 85721

Directional coupling of Thornber-Feynman polarization with the high-$T_c$ ARPES distribution specifies the optimum flatband pseudogap $\Delta$ and mobile localized quasiparticle. This coupling peaks by tuning the statistics and interaction energy to produce stable short-range directional pairing that reflects the lattice asymmetry. Analogous energy gap and BCS ratio parameters are identified for low-$T_c$ long-range acoustical phonons and for quark-anti-quark tightly bound chiral pions in particle physics.

I. Introduction

The pairing mechanism in high-$T_c$ superconductors has remained an elusive entity. A polaron description is suggested by the polarizability of the cuprate lattice, with an off-plane Frohlich coupling constant $\alpha$ of $\approx 5$, compared to 2.5 to 4 in the alkali halides [1]. Further recommending this approach, the open structure aids such a strong long-range interaction, with the observed interaction largely off the plane where screening is minimized. These are arguably the determining system properties and have motivated numerous large and small polaron and bipolaron high-$T_c$ characterizations [2-4]. Inadequately treated however is the low structure symmetry. This should govern any high-$T_c$ polaron state given the nonlocality of the Frohlich self-interaction (induced polarization). In particular, the latter is an extended cumulative process [5] that "sees" this asymmetry as reflected in the high-$T_c$ state, e.g., the flat band (heavy, localized quasiparticle), the short pairing range (~10 è coherence length), and unusual stability (high "boson temperature").

Approaches based on the large or small polaron typically have a problem predicting the mobile nondispersive high-$T_c$ quasiparticle-state. The large polaron is a dispersive negative-energy state that represents the symmetric-structure quasiparticle [6], whereas the small polaron is a tight-binding particle normally associated with the insulating state. In both instances structure geometry is omitted [1]. These factors suggest a different high-$T_c$ polaron approach founded in the interplay of the self-interaction and the geometry. The present paper extends the large polaron approach by incorporating this asymmetry in an effort to better represent the high-$T_c$ system.

In this light a polaron mobility theory due to Thornber and Feynman (TF) explicitly treats low spatial symmetry through a path integral lattice-field balance [7]. Electron recoil motion due to the continuous non-local self-interaction and scattering is tracked in the field by the path integral; a harmonic-coupling approximation for the self-interaction allows it to detail the evolving intermediate state [7]. This mobility quantitatively fits polar data when symmetry is low, e.g., from the field drift at low-temperature, or with negligible drift in a low-symmetry oxide lattice [8,9] (see Appendix A). These considerations recommend a TF characterization for the off-plane interaction in the cuprates, given the short range of the high-$T_c$ state that might accommodate such a one-particle treatment. In addition the governing lattice mode identified from neutron resonance scattering [10] is a shorter range in-plane polar mode more suited to the harmonic approximation.

At present the self-interaction and TF state are not specified in this path/coordinate formulation. However a weak-coupling TF field balance in Appendix A yields a "universal" interaction containing the strong-coupling physics that can simulate polar coupling in the high-$T_c$ valence band system. This model TF interaction resonates with the hole distribution in the $(\pi,0)$ direction defined by angle-resolved photoemission spectroscopy (ARPES) in near-optimum doped $Bi_2Sr_2CaCu_2O_8$ (Bi-2212). The resonance slows the holes to produce the flatband energy shift $\Delta$ as shown in Sect. II.A. The resonance originates with the asymmetry because in that case the self-interaction becomes conserved, so it couples to the dispersive distribution in a *specific direction*. Anisotropic coupling thus dictated defines the energy gap and is consistent with the pairing wavefunction.

The heavy stable quasiparticle and flatband shift/gap have precedents in low-$T_c$ superconductivity and in particle physics in a massive charge-density-wave (CDW) gap "amplitude mode" and gap parameter for the scalar meson mass, respectively, as shown in Sects. II.B and II.C. This short-range Frohlich high-$T_c$ pairing is related to long-range acoustic phonon pairing through common optimum-structure BCS relations in Sects. III.A and III.B. A reformulation of the BCS ratio as an analytic expression in Sect. III.B leads to a further analog in particle physics in the pion-$\bar{q}q$ pair in Sect. III.C, through the quark-level Goldberger-Treiman relation (GTR) and a published derivation of the pion-quark coupling constant. The BCS universality stems from linear dispersion that underlies each of the paired systems. Conclusions from these results are then drawn in Sect. IV.

**II. Energy gap parameter in superconductivity and particle physics**

**A. High-$T_c$ gap from Frohlich coupling to the hole distribution**

Taking the above TF model for the off-plane polarization interaction from the arguments presented, the flatband pseudogap and gap at $(\pi,0)$ specified by the ARPES data in Fig. 1 follow from a resonance coupling to the hole distribution. The gap is produced by an energy shift in the broad distribution peak below the Fermi surface (FS) that defines the band state dispersion. In particular, the peak starts out at large negative energy (~300 meV) below the FS in the $(\pi,\pi)$ direction where it disperses upward. When it reaches negative 40 meV, near $(\pi,0)$, the holes are slowed by a new interaction that stops the dispersion and defines the pseudogap, $\Delta$, as Fig. 1 shows. By coincidence the TF interaction is an energy step that *doubles* $\Delta$ at

≈80 meV, so it couples to the distribution at this point through the occupancies and interaction energy [11]. The step originates with the lattice asymmetry, because it activates the self-interaction to a conserved irreversible process involving the 80 meV in-plane longitudinal-optical (LO) mode (Appendix A and below).

Alternatively, the 80 meV step (from the peak) *straddles* the FS when the peak reaches 40 meV to allow the interaction. Filled and empty initial and final mirror-state occupancies at this point "tune" the interaction to the distribution as illustrated by the ARPES data at (π,0) in Fig. 2. The allowed interaction stops the dispersion to produce the flatband shift and heavy mobile non-wave mechanical particle that derive from the asymmetry. Away from (π,0) the peak disperses from the FS, where the mirror occupancies and distribution width in Fig. 1 exceed the step so the statistics suppresses or decouples the interaction. This directional tuning through the band state dispersion defines an anisotropic coupling.

The LO mode interaction step, $E_{LO}$, is specified by a lattice-field balance for the weak-coupling limit TF mobility, Eq. A-3, where comparison to the Boltzmann mobility shows it also represents the *strong-coupling* balance (Appendix A). This is because the step and a Maxwellian drift distribution imposed together in the balance simulate the drift-asymmetry effect on the self-interaction and distribution in this free-particle limit (see discussion following Eq. A-7; also Ref. 1). For the present purpose the balance also applies in a low-symmetry oxide (Appendix A).

Directional coupling of this interaction to the distribution thereby defines a fundamental relation for the energy-doubled shift 2Δ in the optimum direction as

$$2\Delta = E_{LO}. \qquad (1)$$

The identified resonance LO mode [10] adapts to the TF characterization as noted in Sect. I, and the mode energy ≈80 meV (Table I) gives the predicted doubling in Eq. 1. *Optimum doping* is similarly defined by this coupling criterion, as the doping moves the distribution to where its width again equals or exceeds the interaction step.

Short-range stable pairing follows from the slow dispersion in the resonance region as the slowing bunches the holes spatially. The pairing is a secondary effect of the bunching and without any particle-particle (exchange) interaction: this is reflected in the ultra-short lifetime and range of the state. A d-wave pairing wavefunction mirrors the above directional coupling.

The shift/pair stability (high boson temperature) follows from the conserved high-energy interaction, which again manifests the asymmetry as developed in Appendix A. This inelastic self-interaction contrasts with the off-shell large polaron interaction [6,5]. More to the point, the TF-derived state conforms with the short-range stable high-$T_c$ state as discussed here.

The self-interaction step additionally equals the single-particle energy shift: the latter is gleaned from a decomposition of the TF mobility expression [9]. This shift accordingly predicts the gap 2Δ *independent* of the distribution and FS, in harmony with the short-range state.

Table I lists the high-$T_c$ gap (2Δ) and interaction energy/energy shift ($E_{LO}$) for the various structures, with the data updated from a similar listing in Ref. 1. The shift-gap match in the diverse structures indicates a common polar lattice origin, e.g., the factor six scaling from the cuprates to $Rb_2CsC_{60}$.

## B. Nondispersive gap in the CDW in low-$T_c$ superconductors

The preceding flatband high-$T_c$ state near $(\pi,0)$ has a long-range precedent in the CDW in the low-$T_c$ superconductors, where a flatband gap related to the superconducting gap ($2\Delta$) is created by lattice distortion of the FS. In particular, optical phonon-induced oscillation of the normal-state CDW gap creates a massive collective mode called an "amplitude mode" (AM) that couples to $2\Delta$, as shown by Raman scattering in $NbSe_2$ [22]. The mode is common to the low-$T_c$'s and is produced from a time-dependent perturbation of the FS due to the lattice distortion as portrayed in Fig. 3; a Bragg-reflection plane (through the FS) induced by the distortion opens the gap, shown in the inset, with the result the band is locally flattened. The CDW-AM gap is 40 cm$^{-1}$ compared to a $2\Delta$ of 17 cm$^{-1}$, and the coupling transfers spectral weight between them as shown in Fig. 4. (The coupling in the high-$T_c$'s in contrast is *direct* as expressed by Eq. 1, because the band state disperses and "tunes" to the polarization energy as discussed in Sect. II.A). The flatband CDW-AM at $4.6\Delta$ in Fig. 3 is thus evidently the long-range analog of the normal high-$T_c$ state in Fig. 1, where the mass generation in the latter is from the broken symmetry.

The CDM-AM is identified in Ref. 22a with a $C^+$ (charge-conjugation "even") electron-hole bound state close to $2\Delta$ with "mass" $M_{C+}$,

$$2\Delta = M_{C+}, \qquad (2)$$

which parallels the Nambu scalar mass that predicts the sigma meson mass [22b] and is discussed further in Sect. II.C.  This contrasts with the massless $C^-$ (charge-conjugation "odd") electron-hole bound state [22,23], which parallels the Nambu-Goldstone (massless) pion [24].

### C.  Gap parameter for the sigma meson

The independent specifications of the flatband gap and massive CDW gap amplitude mode in high- and low-$T_c$ superconductivity respectively in Sects. II.A and II.B have an analog in particle physics in a gap parameter, the Nambu scalar meson mass [25]

$$2m_q = m_S, \qquad (3)$$

where $m_q$ is the constituent quark mass.  This was also concluded in Ref. 22, where the CDW-AM was related to a chiral-breaking analog mode of Nambu and Jona-Lasinio [26].  Eq. 3 is verified experimentally by the sigma meson, where $m_\sigma \approx 600$ MeV [27] and an independent estimation for the quark mass is $m_N/3 \approx 315$ MeV (in units c=1).  Furthermore $m_q$ is derived in the chiral limit (CL) for the linear $\sigma$ model as 325 MeV and $m_\sigma$ as 650 MeV [28].  Equation 3 is an expression of the high-mass limit in the dispersion relation range 0 to $4m^2$, and represents an energy region where the quark-anti-quark pair can be considered as "touching".  The low-mass limit taken in Sect. III.C yields a tight-binding limit "fused" $\bar{q}q$ pair for the pion.

### III.  BCS ratio parameter in superconductivity and particle physics

## A. Optimum high-$T_c$ gap ratio and lattice structure

The high-$T_c$ gap specified by the conserved self-interaction energy $E_{LO}$ in Eq. 1 leads to fundamental structure criteria needed to maximize the critical temperature $T_c$. This temperature is given by the predicted gap divided by the minimum BCS gap ratio (g.r.) 3.5:

$$T_c = 2\Delta/(k_B \times \text{g.r.}) \rightarrow E_{LO}/(3.5 k_B) \approx 250° \text{ K}. \qquad (4)$$

The prediction $2\Delta$ was verified in Table I with the polar mode energies $E_{LO}$, and it was noted the latter is also the quasiparticle energy shift. The BCS ratio in the denominator is obtained in structures that optimize the interaction from nearby uniformly-spaced polar modes [1]. Cubic structures satisfy this criterion, and the experimental ratios in cubic $Rb_2CsC_{60}$ and $BaKBiO_3$ are in fact 3 to 4 (Table I), averaging to 3.5.

Note that higher $T_c$ scattering from the optimum interaction and (BCS) gap ratio in Eq. 4 offsets the polarization, which works against the $T_c$ increase consistent with the gap ratio-$T_c$ proportionality in Table I. This counter-interaction also flattens the polar mobility data fit by the TF mobility [8,9]; the dual effect in the latter framework is detailed in Appendix A. Scattering offset is further evident in high-$T_c$ superconductivity in the inverse-$T_c$ variation of the isotrope effect. The offset however does not restrict the predicted $T_c$ when there are sufficient polar modes as the source of the self-interaction: hence the optimum parameters on the righthand side of Eq. 4 are *simultaneously* achieved in cubic alkali-$C_{60}$ (see Table I). In this case a low lattice frequency restricts $T_c$.

An optimum cuprate structure with effective cubic symmetry and $T_c$ given by Eq. 4 can be obtained by adding planes and polarization for the added planes, specifically the long-range ferroelectric (FE) c-axis Cu-O mode. This mode addition is indicated by the plane-number dependence in which $T_c$ decreases after three planes and the c-axis mode is only on the outer plane. The short-range 80 meV ab-plane LO mode identified with $2\Delta$ in Sect. II.A is present with each plane. These considerations indicate the c-axis mode starts a two-step nonlinear interaction, and this view is supported by nonlinearities observed in the mode couplings [29].

The above consideration of the symmetry and the polarization sources indicate the (250° K) limit can be realized by adding FE c-axis modes and doping for the inner planes of an infinite-layer structure. One approach is to add oxygen (O) and an oversize dopant to the cation layer; this will also increase the plane spacing, to restore cubic (perovskite) symmetry, and is discussed further in Ref. 1.

Importantly for the relation to BCS pairing, the global dispersion is linear due to the large FS [30], which follows from the constant of proportionality defined from energy-momentum changes $\Delta E$ and $\Delta p$,

$$\Delta E = (p/m) \Delta p. \qquad (5)$$

Here the factor $p/m$ is approximately constant on the large FS because the relative changes, $\Delta p/p$, are small [31]. The *local* momentum on the FS however is not well-defined, i.e., on the 0.1 eV interaction-energy scale that specifies the quasiparticle in Sect. II.A. This is because of irreversibility in the continuous path-integral interaction in Ref. 7, as modeled in Appendix A.

**B.  Long range-low energy BCS pairing**

The short-range Frohlich pairing in high-$T_c$ superconductivity then relates to long-range acoustic phonon pairing through the analogous high-mass flatband states that define $\Delta$ in each case (Sects. II. A and II.B), and common BCS gap ratio and linear dispersion. For the acoustic phonon case the linearity follows from the coupling; for low-energy acoustic modes with frequency $\omega$ and wavenumber q,

$$\omega \sim c_s q, \qquad (6)$$

where the proportionality constant $c_s$ is the speed of sound.

An analytic expression derived for the BCS ratio [32] suggests these pairing limits in condensed matter physics are related by analogy to the pion-$\bar{q}q$ pair in particle physics in the opposite limits of loose- and tight-binding. In particular the BCS expression is a ratio of energy-dimension quantities that characterize the system, and it is evaluated as a manifest dimensionless number,

$$2\Delta/k_B T_c = 2\pi e^{-\gamma} = 3.528, \qquad (7)$$

where $\gamma$ is the Euler constant (0.5771) and the derivation from Ref. 32 is given in Appendix B. The (3.528) magnitude proportional to $2\pi$ in Eq. 7 provides the indicated connection of BCS pairing to the pairing in particle physics to be discussed next.

**C. Tight binding limit in particle physics; chiral $\bar{q}q$ pair**

The pairing cases in condensed matter physics just presented further relate to pion-$\bar{q}q$ pairing in particle physics through a BCS-like ratio obtained from a quark-level Goldberger-Treiman relation (GTR) and a published derivation of the pion-quark coupling constant. In particular, the GTR is the well-known characteristic equation for the pion-nucleon system, and expresses the pion-nucleon coupling constant, $g_{\pi NN}$, as a dimensionless ratio of the nucleon mass ($m_N$) times the axial current form factor, $g_A$, divided by the pion decay constant, $f_\pi$, i.e.,

$$g_{\pi NN} \approx m_N g_A / f_\pi. \qquad (8)$$

Here $g_{\pi NN}$ is experimentally determined to be ~13.4, $f_\pi$ is ~90 MeV (million electron volts) in the CL [28], and $g_A$ is 1.26, where the latter reflects the three-quark nucleon structure. An equivalent (dimensionless) GTR relation can be written in terms of the constituent quark mass ($m_q$) and pion-quark coupling constant ($g_{\pi \bar{q} q}$) as

$$g_{\pi \bar{q} q} \approx m_q / f_\pi, \qquad (9)$$

where the constituent quark mass $m_q$ is $\approx$320 MeV as in Eq. 3 and the axial form factor ($g_A$) does not enter for the structureless quark [33]. Hence an experimental-based $\pi \bar{q} q$ coupling constant $g_{\pi \bar{q} q}$ is 320 MeV/90 MeV ~ 3.6, where analogy of Eq. 9 to the BCS relation (Eq. 7) is apparent in the magnitude of the dimensionless ratio.

Noting that the righthand side of Eq. 9 is a dimensionless ratio of characteristic energies due to axial current conservation, $\partial . j^A = 0$ (corresponding to a massless Goldstone pion), the above connection to BCS pairing is pursued by writing Eq. 9 in "BCS form", from which an analog "critical temperature" and quark pair binding are defined. Multiplying the numerator and denominator by two,

$$g_{\pi \bar{q} q} = 2 m_q / 2 f_\pi, \qquad (10)$$

BCS analogs from Eq. 7 are $2\Delta \rightarrow 2 m_q \approx 640$ MeV for the pair binding energy, and $T_c \rightarrow 2 f_\pi$ ~180 MeV (in units $k_B = 1$) for the effective critical temperature. Moreover $2 f_\pi$ is independently derived as the critical temperature for creation of the paired state [34], i.e., a "chiral restoration temperature". A BCS-like relation accordingly follows from Eq. 10 as

$$g_{\pi\bar{q}q} = 2m_q/T_c. \qquad (11)$$

This connection is supported by an analytic expression derived for $g_{\pi\bar{q}q}$, where both the *magnitude and form* are similar to that of the BCS ratio (Eq. 7). The expression for $g_{\pi\bar{q}q}$ as a manifestly dimensionless number is obtained in the linear σ model from a nonperturbative bootstrap as [28]

$$g_{\pi\bar{q}q} = 2\pi/3^{1/2} \approx 3.628, \qquad (12)$$

with near-identical magnitude and $2\pi$ factor in common with the (manifest dimensionless) BCS ratio, Eq. 7.

The lightness of the pion at 140 MeV compared to the binding ($2m_q$) indicates the pair is tightly bound, with the quark mass "tied up" in the binding energy (large mass defect), so the pair can be regarded as "fused". Tight binding is further implied from the GTR, from the Goldstone (zero pion mass) limit taken for axial current conservation, e.g., the quark mass is *totally* taken up in the binding. The chirality of the paired state similarly follows in the massless limit.

This tight-binding Goldstone fusion limit is solidified from an independent theoretical derivation and measurement of the pion charge radius, $r_\pi$ [35],

$$r_\pi = 3^{1/2}/(2\pi f_\pi) \qquad (13)$$

in units where $\hbar=c=1$, which with Eqs. 12 and 9 give in the CL

$$r_\pi = 1/m_q = 0.61 \text{ fm}, \qquad (14)$$

where $m_q$ is taken from the GTR $m_q \approx f_\pi g_{\pi\bar{q}q} = 325$ MeV for $f_\pi \sim 90$ MeV in the CL. This compares to the experimental charge radius $0.63 \pm 0.01$ fm [36]. The radius determined by the *single* quark mass again indicates the pair binding is extremely tight, where the quark and anti-quark blend or fuse together.

Linear dispersion again follows, from the relativistic limit taken in the GTR for axial current conservation. Hence $E^2=m^2+p^2 \to p^2$, i.e., $E \to p$, the limiting dispersion is linear, and the pion quark-anti-quark pair constitutes a pure Goldstone pion state.

**IV. Conclusions**

We have shown that Frohlich polarization, represented by a strong-coupling model Thornber-Feynman (TF) interaction, directionally couples to the ARPES distribution in Bi-2212. This coupling specifies the optimum flatband high-$T_c$ pseudogap $\Delta$ through the resonance relation Eq. 1, as described in more detail in Sect. II.A. Resonance occurs when the distribution disperses to $(\pi,0)$ because the Fermi surface (FS) lies midway between the initial/peak and final energies in this direction, so the statistics allows the interaction. Extension of the single-particle TF theory to the cuprates is based on its applicability in low-symmetry polarizable media and the short range of the high-$T_c$ state. The interaction is derived in Appendix A as an irreversible energy step $E_{LO} \approx 80$ meV due to the asymmetry that doubles the 40 meV peak-to-FS energy at $(\pi,0)$, to naturally define the optimum/resonance direction and doping. The allowed interaction slows the holes to create the flatband shift/gap and mobile non-wave mechanical quasiparticle that reflect the asymmetry. Minimal stable pairing consistent with d-wave pairing follows from the slowing. The stability (high boson temperature) and localization (heavy mass) derive from the energetic self-interaction that again originates with the asymmetry.

The above massive Frohlich quasiparticle with short-range stable high-$T_c$ pairing have analogs in long-range low-$T_c$ Cooper pairing and also in tight-binding quark-anti-quark $\bar{q}q$ pairing through gap and BCS ratio parameters that characterize each, as summarized in Sects. II.B, II.C and III. Nonrelativistic and relativistic limits yield these analog relations for the sigma meson and pion in Eqs. 3 and 11, respectively. This common BCS ratio is elucidated by a published dimensionless analytic expression for the ratio and also for the pion-quark coupling constant. The latter constitutes an analog "BCS ratio" identified from the quark-level Goldberger-Treiman relation (GTR); an alternate derivation of the BCS ratio expression is given in Appendix B. Near-equal BCS-like ratios of 3.528 in Eq. 7, 3.5 in Eq. 4 (from cubic structure data in Table I), and 3.628 in Eq. 12 are not accidental. This numerical universality is founded in common linear dispersion with energy proportional to momentum that fundamentally links the pairing in the binding limits of condensed matter and particle physics.

**Appendix A: Polarization interaction in low-symmetry systems**

Asymmetry energizes the self-interaction to a conserved irreversible process displayed here in a low-order lattice-field momentum balance from Ref. 7. The balance gives the weak-coupling limit of the TF mobility, Eq. A-7 with $\alpha \to 0$, but as noted it also models the polarization physics. This is evident from the coupling-independent temperature ratio $(3/2)T/T_{LO}$ in the mobility expression which distinguishes it from Boltzmann polaron mobilities, and this is discussed further following Eq. A-7. Low symmetry in cubic polar lattices occurs at low-tempertaure from long field-drift: extension to the cuprates is based on a wide-range TF mobility fit in tetragonal $TiO_2$ where drift is nullified in the upper temperature range [9]. This low-order balance specifies the interaction kinematics that couples to the hole distribution in Sect. II.A.

The lattice rate for the balance is the integrated difference of induced Frohlich emission and thermal scattering (absorption) times the phonon momentum **q**, where conservation is imposed by the Fermi golden rule (cf. Eq. 16 in Ref. 7). (The Frohlich potential is left exact for these free-particle rates as it is for the nonlocal scattering rate in the path integral calculation; see Sect. I). Weighting the rates with the local distribution f(**p**′), the balance is

$$-e\mathbf{E} = 2\pi\Sigma |C_q|^2 \mathbf{q} \int d^3\mathbf{p}' f(\mathbf{p}') \{ N_{LO} \delta[\omega_{LO} - \mathbf{q}\cdot\mathbf{p}/m_b - q^2/2m_b] +$$

$$- (N_{LO}+1) \delta[-\omega_{LO} + \mathbf{q}\cdot\mathbf{p}/m_b - q^2/2m_b]\}, \quad (A\text{-}1)$$

in terms of the applied field **E** and unit phonon number density $N_{LO} = [\exp(\omega_{LO}/T)-1]^{-1}$, where $\omega_{LO}$ is the LO mode frequency. Other parameters are the band mass $m_b$, Frohlich amplitude $C_q = -(\omega_{LO}/q)(2m_b\omega_{LO})^{-1/4}(4\pi\alpha)^{1/2}$, and local and total momenta **p**′ and **p**, respectively, where **p**′=**p**-$m_b$**v**, **v** is the drift velocity, and =$k_B$=1 in the units used. This distribution f(**p**) was anticipated in Ref. 7 as a drifted Maxwellian $f_0(\mathbf{p}'+m_b\mathbf{v})$ because in this case the free-particle limits of the path integral and low-order TF field balances (Eqs. 13 and 17 in Ref. 7) are identical. So the local distribution in Eq. A-1 is taken as

$$f_0(\mathbf{p}') = (\beta/2\pi m_b)^{3/2} \exp[-\beta(\mathbf{p}')^2/2m_b], \quad (A\text{-}2)$$

where $\beta$ is the inverse temperature. Then the balance Eq. A-1 becomes

$$-e\mathbf{E} = 2\pi(\beta/2\pi m_b)^{3/2} \Sigma |C_q|^2 \mathbf{q} \int d^3\mathbf{p}' \exp[-\beta(p'^2/2m_b)] \{(N_{LO}+1)$$

$$\times\delta[-\omega_{LO} + \mathbf{q}(\mathbf{p}'+m_b\mathbf{v})/m_b - (q^2/2m_b)] - N_{LO} \delta[\omega_{LO} - \mathbf{q}(\mathbf{p}'+m_b\mathbf{v})/m_b - (q^2/2m_b)]\}. \quad (A\text{-}3)$$

Note the cross term $\mathbf{q}(\mathbf{p}'+m_b\mathbf{v})/m_b$ expresses local momentum conservation about the drift velocity $\mathbf{v}$.

Equation A-3 contains the asymmetry (the drift $\mathbf{v}$) and associated irreversibility (the energy step $\omega_{LO}$) used to couple to the distribution in Sect. II.A, so it remains to connect this equation to the TF self-interaction. This is provided by the above weak-coupling TF balance, Eq. 17 in Ref. 7, because it is the Fourier time transform of Eq. A-3. Recall Eq. 17 is

$$eE = \int dt \sum |C_q|^2 \mathbf{q} \{ (N_{LO}+1) \exp[it(\omega_{LO} - \mathbf{v}\cdot\mathbf{q})] - N_{LO}\exp[-it(\omega_{LO} - \mathbf{v}\cdot\mathbf{q})]\}$$

$$\times \exp\{-(q^2/2m_b)[-it + (t^2/\beta)]\}, \qquad (A-4)$$

where the transform relationship to Eq. A-3 is shown through the time and momentum integrals

$$\delta(a) = (2\pi)^{-1} \int dt\, e^{-iat}, \qquad (A-5)$$

and [37]

$$(\beta/2\pi)^{3/2} \int d^3\mathbf{p}' \exp[-\beta(p^2/2) + it\mathbf{p}'\cdot\mathbf{q}] = \exp(-q^2t^2/2\beta). \qquad (A-6)$$

(Note the Frohlich amplitude in Eq. A-4 is Doppler-shifted due to the drift, with $e^{i\omega t} \to e^{i(\omega - \mathbf{q}\cdot\mathbf{v})t}$). The weak-field limit of Eq. A-4 (velocity per unit field as $\mathbf{E}\to 0$) then gives the $\alpha\to 0$ limit TF mobility, where the full TF mobility is [7]

$$\mu_{TF} = (3/2)(T/T_{LO})\mu_B. \qquad (A-7)$$

Here $\mu_B$ is the large-coupling Boltzmann polaron mobility [38] and $T_{LO}$ is the effective lattice "temperature" $E_{LO}$.

Importantly the integer ratio $(3/2)T/T_{LO}$ in the mobility contains the noted irreversible polarization effect even in this free-particle limit. This is

because.conservation (delta-function kinematics) and the Maxwellian drift distribution imposed in Eq. A-3 are *mutually consistent* and simulate the drift effect. Hence drift activates the self-interaction via the irreversible step $\omega_{LO}$, which acts back against the field and maintains local equilibrium. Alternatively, the existence of the latter evidences its vigorous restoration against the field by the conserved interaction.

      The lattice counteraction in Eqs. A-1, A-3, and A-4 appears in the mobility *and data* as a temperature flattening relative to Boltzmann polaron mobilities [8,9], again through the temperature ratio in Eq. A-7. This flattening at low temperature evidences dominance by the self-interaction in this range, so the mobility dips below a Boltzmann mobility. (A self-interaction factor in the ratio taken with the corresponding Boltzmann-scattering factor quantifies this dual lattice effect in Ref. 9). With hindsight, the gap ratio-$T_c$ variation in Table I and inverse-$T_c$ isotope effect variation in high-$T_c$ superconductivity also manifest the above scattering counter-interaction as noted in Sect. III.A.

**Appendix B: Analytic expression for the BCS ratio**

Defining equations for the energy gap $\Delta$ and BCS parameter $N(0)V$ are, from Eqs. 2-34 and 2-78 in Ref. 39,

$$\Delta = 2\omega e^{-1/N(0)V} \qquad (B\text{-}1)$$

$$1/N(0)V = \int d\varepsilon\, \varepsilon^{-1} \tanh(\varepsilon/2k_B T_c) \quad \text{(limits 0 to } \omega\text{)} \qquad (B\text{-}2)$$

$$= \log x \tanh x \Big| - \int dx \log x\, \text{sech}^2 x \quad \text{(limits 0 to } \omega/2k_B T_c\text{)} \qquad (B\text{-}3)$$

where the second equality is obtained from an integration by parts. Now since $\omega/2k_B T_c \gg 1$, the upper limit of the integral can be replaced by infinity and an integral related to it identified in tables of integrals. From Ref. 40

$$\int dx \log[\log(1/x)]/(1+x)^2 = (1/2)[\log(\pi/2) - \gamma] \quad \text{(limits 0 to 1)} \qquad (B\text{-}4)$$

where $\gamma$ is the Euler constant defined as $-\int e^{-r \log r}\, dr$ (limits 0 to $\infty$), and is numerically equal to 0.5771. Letting $x = e^{-2y}$, so that $\log(1/x) = 2y$, the integral in Eq. B-4 is found to equal

$$(1/2)\int dy\, (\log 2 + \log y)\, \text{sech}^2 y \quad \text{(limits 0 to } \infty\text{)}$$

from which the integral in Eq. B-3 is evaluated as

$$\int dy \log y\, \text{sech}^2 y \cong \log(\pi/4) - \gamma. \qquad (B\text{-}5)$$

Eq. B-3 then becomes

$$1/N(0)V = \log(\omega/2k_B T_c) + \gamma - \log(\pi/4) \qquad (B\text{-}6)$$

which with Eq. B-1 gives for the gap ratio,

$$2\Delta/k_B T_c = 2\pi e^{-\gamma} = 3.52775 \qquad (B\text{-}7)$$

The expression of the ratio in analytic form evidences the stated fundamental property of BCS pairing. (Note the replacement of the upper limit of integration in Eq. B-3 is fairly good for the cuprates also, as $\omega/k_B \sim 1000°$ K for the in-plane polar mode).

**Acknowledgements**

We are indebted to T. Egami for the coupling mechanism to the hole distribution in Sect. II.A, K. K. Thornber for assistance in deriving the mobility field-balance in Appendix A, and R. Parmenter for a derivation of the analytic expression for the BCS ratio in Appendix B.

scattering resonance, evidencing precursor boson-quasiparticle formation. See also Ref. 11.

|  | Optimum gap, $2\Delta$ (meV) | Interaction energy/ polarization shift, $E_{LO}$ (meV) | Gap ratio, $2\Delta/k_BT$ |
|---|---|---|---|
| Bi-2212 | 75 [13,14] | 80 [15,16] | 9 |
| $YBa_2Cu_3O_7$ (YBCO) | 60 | 75 [16,17] <br> $62^a$ [17] | 7 |
| $La_{1.85}Sr_{0.15}CuO_4$ (LSCO) | 15 <br> $90^b$ [18] | 85 [10] | 5 |
| $Rb_2CsC_{60}$ | 12 [19] | 13 [20] | 3.5 |
| $BaKBiO_3$ (BKB) | 9 | 30 [21] | 3-4 |

[a]Shift from the c-axis chain mode.

[b]Pseudogap measured below $T_c$

**TABLE I.** Match of the optimum gap/pseudogap $2\Delta$ to the interaction energy or quasiparticle energy shift ($E_{LO}$) from the low-symmetry Frohlich polarization. Shifts involving the ≈80 meV in-plane polar LO mode are based on neutron resonance scattering in Ref. 10. The factor six prediction range of the match from the cuprates to $Rb_2CsC_{60}$ indicates a common low-symmetry polar description. Small gaps in LSCO and BKB relative to the shift reflect short-range order and heavy doping, respectively: note the match in LSCO to a 90 meV measured pseudogap. Large gap ratios relative to 3.5 reflect a higher $T_c$ scattering offset that flattens the polar mobility.

**FIGURE CAPTIONS**

**Figure 1.** Momentum (k) dependence of the ARPES distribution for near-optimum doped Bi-2212 along the $\Gamma$-M-Z line, $(\pi,\pi)$ to $(\pi,-\pi)$ directions, for the normal and superconducting states, panels (a) and (b), respectively. The peaks define the hole dispersion, where a flatband shift at 40 meV below the FS (zero energy) defines the pseudogap $\Delta$.(from Ref. 12).

**Figure 2.** ARPES distribution of Bi-2212 in Fig. 1 at $(\pi,0)$ that defines $\Delta$, from the 40 meV shift below the FS and empty "mirror-image" final state 40 me'V above, as discussed for the normal and superconducting states (from Ref. 12).

**Figure 3.** Schematic picture of the Fermi surface of 2H-NbSe$_2$, showing a Bragg plane AB induced by lattice distortion of the CDW state that parallels the flatband high-T$_c$ state in Fig. 1 (adapted from Ref. 22a). The inset shows how the FS changes from the normal state (full line) to the CDW state (dashed line), and with the lattice perturbation to the CDW-AM (dotted line).

**Figure 4.** Spectral weight of the CDW-AM depicted in Fig. 3 due to 15% transfer to the superconducting gap at 2$\Delta$, from lattice coupling to the gap (adapted from Ref. 22a).

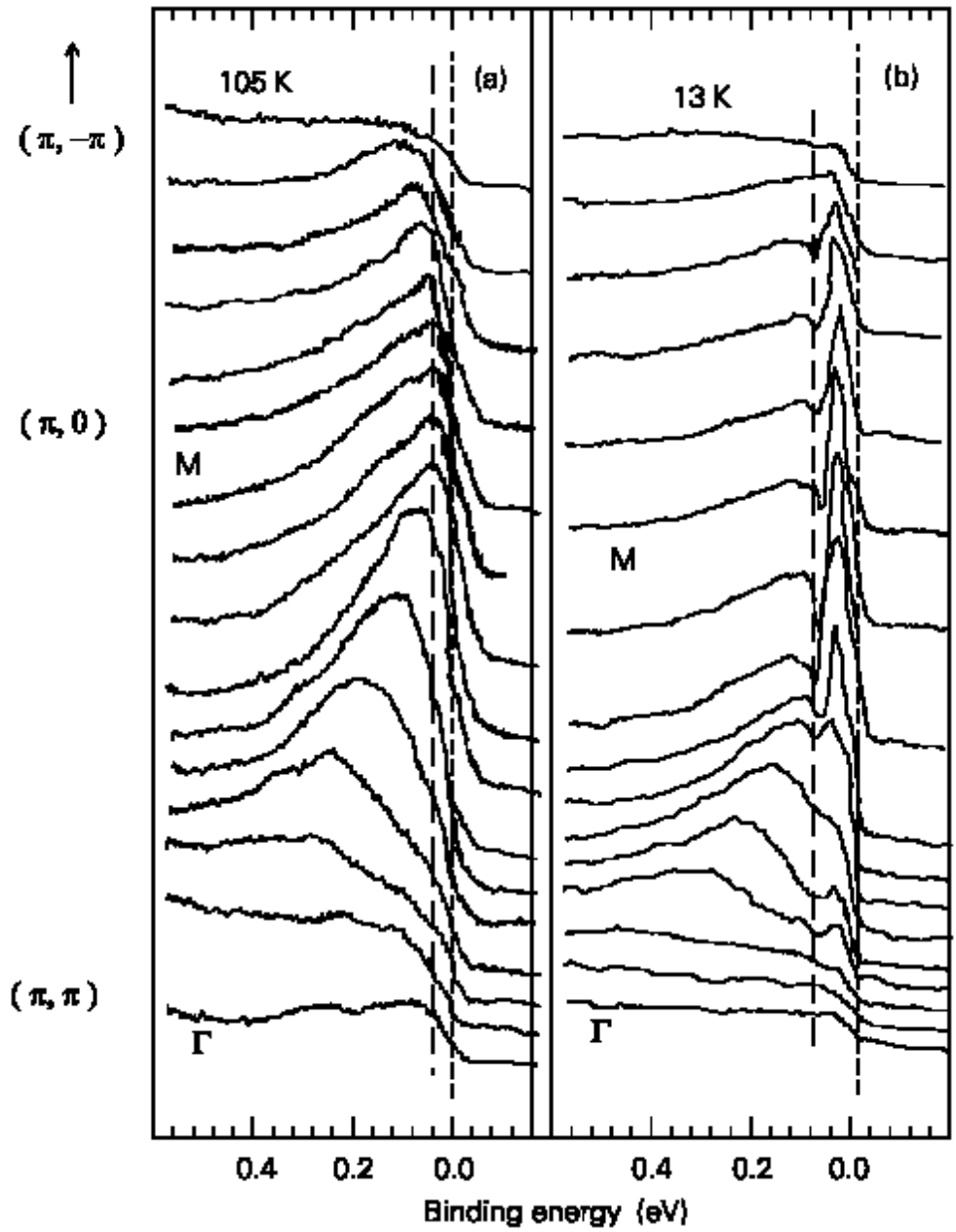

**Figure 1**

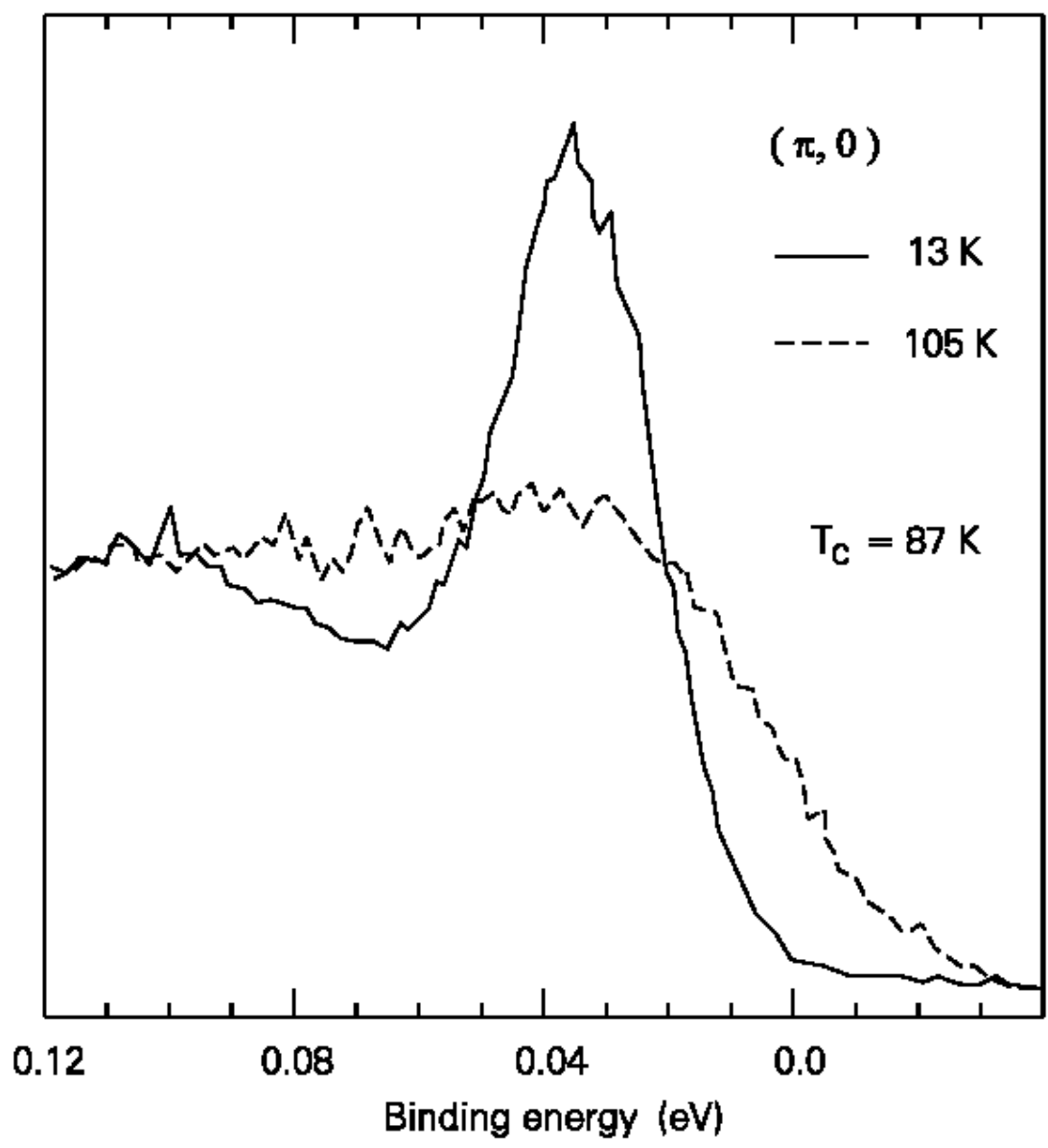

**Figure 2**

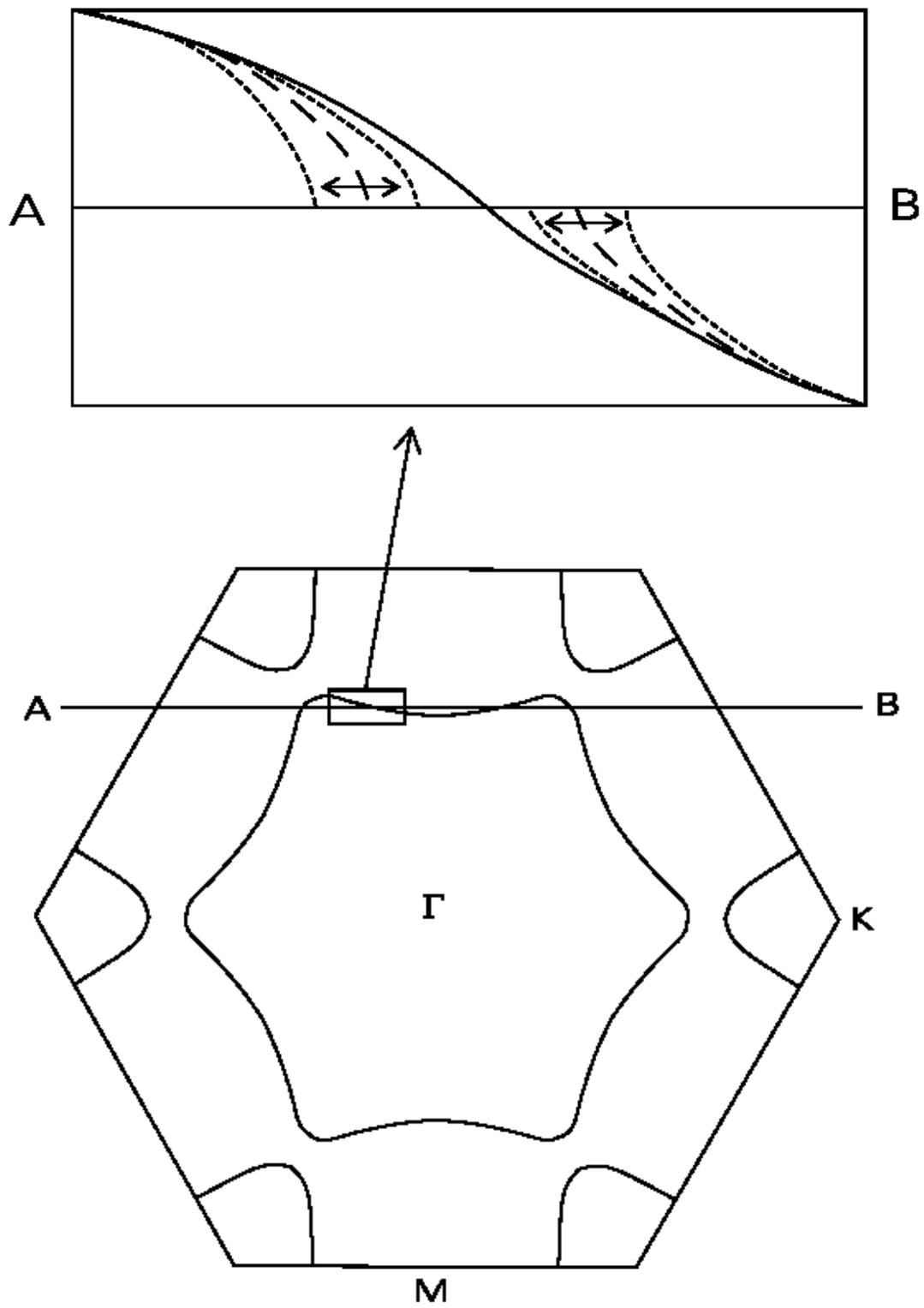

**Figure 3**

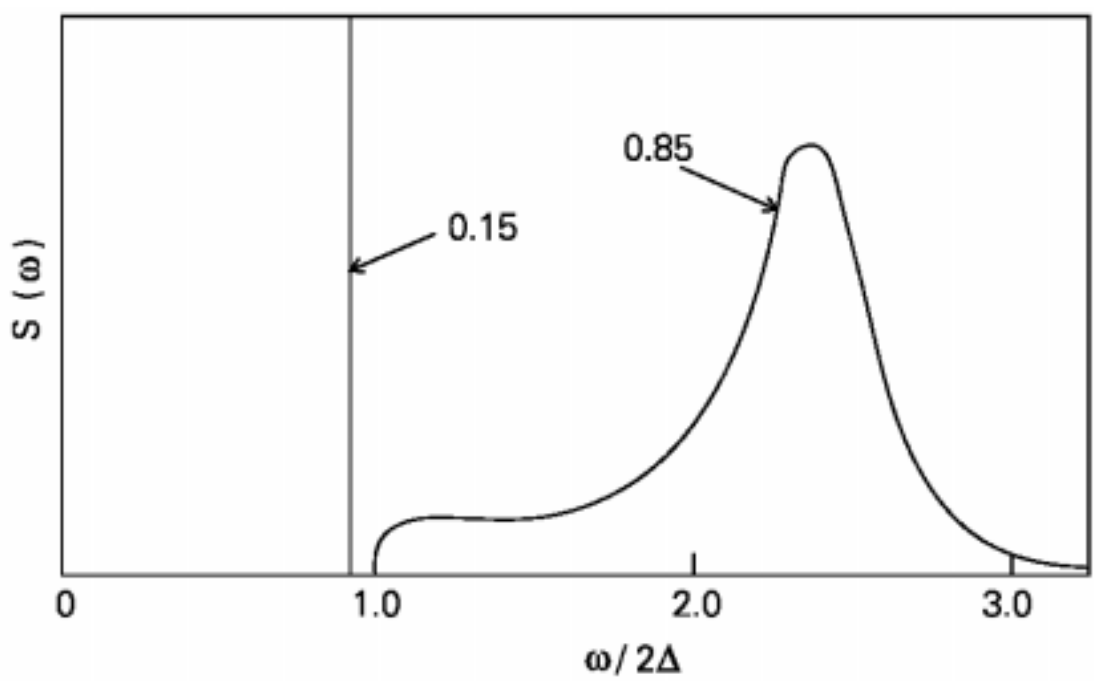

**Figure 4**